\documentclass[11pt]{article}
\newcommand{\writingdate}{July 23rd, 2000}
\input epsf.tex

\newcommand{\room}{\rule[-0.3cm]{0cm}{0.8cm}}

\newcommand{\s}{\sigma}

\newcommand{\mc}{\mathcal}

\newcommand{\bk}{\mbox{\boldmath$k$}}
\newcommand{\bm}{\mbox{\boldmath$m$}}

\newcommand{\bxi}{\mbox{\boldmath$\xi$}}

\newcommand{\bt}{\mbox{\boldmath$T$}}
\newcommand{\bR}{\mbox{\boldmath$R$}}

\newcommand{\bs}{\mbox{\boldmath$\sigma$}}

\title{\LARGE\bf
Attractor Modulation and Proliferation \\
in $1+\infty$ Dimensional Neural Networks}
\date{\writingdate}
\author{\Large N.S.\@ Skantzos \hspace*{15mm} A.C.C.\@ Coolen \\[3mm]
Department of Mathematics, King's College London\\
The Strand, London WC2R 2LS, UK}

\begin{document}

\maketitle
\begin{abstract}
\noindent
We extend a recently introduced class of exactly solvable models
for recurrent neural networks with competition between 1D nearest neighbour
and infinite range information processing.
We increase the potential for further frustration and competition in these
models, as well as their biological relevance, by adding next-nearest neighbour couplings, and we
allow for modulation of the attractors
so that we can interpolate continuously between situations with different numbers
of stored patterns. Our models are solved by combining
mean field and random field techniques. They exhibit increasingly complex phase
diagrams with novel phases, separated by multiple first- and second order transitions
(dynamical and thermodynamic ones), and, upon modulating the attractor
strengths, non-trivial scenarios of phase diagram deformation.
Our predictions are in excellent agreement with
numerical simulations.
\end{abstract}

\begin{center}{
}\end{center}

\tableofcontents

\clearpage
\section{Introduction}

In real (biological) recurrent neural networks, where information processing is
based on the creation and manipulation of attractors, one typically observes
an intricate interplay and competition between long-range information
processing (via excitatory pyramidal neurons) and short-range information
processing (via short-range pyramidal neurons and inhibitory
inter-neurons). Studying those properties of such systems which are linked to
their spatial structure, using
statistical mechanical techniques, requires moving away from the more traditional
 infinite range models of attractor neural networks \cite{hopfield,amit}.
With the latter objective, an alternative type
of attractor neural networks was recently proposed and studied \cite{skantzoscoolen},
in which neurons (represented by Ising spins)
are mutually connected by a combination of infinite range synaptic  interactions, and
 one-dimensional nearest-neighbour interactions. Although real biological network
 architectures  are obviously far more complex, such models,
which are still sufficiently simple to be solved exactly, via a combination of  mean-field
techniques (as in e.g. \cite{amit}) and random field
techniques (as in e.g. \cite{bruinsma,rujan,mukamel}), would appear to represent a
small but welcome step towards biological reality.
Moreover, from a statistical mechanical perspective,
the solutions of these models  exhibited a remarkably rich
behaviour, even in the so-called low storage regime (where the number of patterns stored in the interactions
remains finite in the thermodynamic limit), and
particularly in those regions in parameter space where the two
types of interactions (long range versus nearest-neighbour)
compete most strongly. The phase diagrams where found to describe a series of
regions with different numbers of ergodic components, separated
by both second and first order transitions (representing various dynamical transitions, in addition
to the thermodynamic ones), and to increase dramatically in complexity
with the  number $p$ of stored patterns.

The present paper is devoted to a further exploration and enlargement
of the class of models introduced in \cite{skantzoscoolen}.
We study two orthogonal extensions, each with their own specific
objectives. Our first extension is to include also neuronal interactions between
 next-nearest neighbours in the 1D-chain (in addition to the mean-field and nearest neighbour ones),
 and to study their impact on the phase
 diagrams.
Here the motivation is, again, partly biological: short-range
pyramidal neurons are believed to act on shorter distances than
 short-range inter-neurons, and simple models of the type proposed here
 have indeed been used recently to explain properties of the
 mammalian visual system \cite{sompolinsky,douglas}.
We find, especially when the parameters controlling the new interactions are chosen such as to
introduce further
competition and frustration into the network, new phases are being
created and the complexity of the phase diagram is again significantly
increased.
Our second extension is primarily motivated by our desire to understand the
significant qualitative modification of the phase diagrams as observed in \cite{skantzoscoolen}
resulting from just a small increase in the number of stored patterns
(e.g. $p=1$ versus $p=2$).
More specifically, in contrast to the traditional long-range
models, in models with short-range interactions one finds a
stronger disruptive effect of non-condensed patterns on the recall of the
condensed ones.
In order to shed light on such phenomena we extend the
original models of \cite{skantzoscoolen} by modulating the
embedding strengths of the individual stored patterns (and therefore the attractors themselves),
as in \cite{viana},
so that we can smoothly interpolate between, for instance, the $p=1$
and $p=2$ models. This reveals, as was expected on the basis of the qualitative
differences between the $p=1$ and $p=2$ diagrams, a very complicated
and interesting scenario of phase diagram deformation.
Both extensions of the original models in \cite{skantzoscoolen}
introduce technical complications, but these are largely of a
quantitative nature, and the extra work needed to again arrive at exact
solutions is more than  adequately compensated by the richness of the resultant
phase diagrams.

\section{Definitions}

As in \cite{skantzoscoolen},
each of our extended models is defined as a collection of $N$ binary  neuron
variables (i.e. Ising spins) $\bs=(\s_1,\ldots,\s_N)$, with $\s_i\in\{-1,1\}$, which
evolve in time stochastically and sequentially, following the Glauber-type rule
\begin{equation}
{\rm Prob}\left[\s_{i}(t+
1)=\pm 1\right]=\frac12\left[\room
1\pm\tanh[\beta h_i(\bs(t))]\right]
\hspace{13mm}
h_{i}(\bs)=\sum_{j\neq i}J_{ij}\s_{j}
\label{eq:dynamics}
\end{equation}
The parameters $J_{ij}$ represent the synaptic interactions, and the
parameter $\beta=1/T$ controls the amount of stochasticity in the
dynamics. If the interaction matrix is
symmetric, the process (\ref{eq:dynamics}) leads to a unique
equilibrium state of the Boltzmann type, i.e. with microscopic state
probabilities of the form $p_\infty(\bs)\sim \exp[-\beta H(\bs)]$ and with the
conventional Ising  Hamiltonian $H(\bs)=-\sum_{i< j}\s_i J_{ij}\s_j$.
Information processing in such systems is based on the creation
and manipulation of attractors in the system's configuration
space, by a suitable choice of the spin-interactions $\{J_{ij}\}$, which shape the energy landscape.
In
statistical mechanical terms, the  two key aspects of these
interactions which determine the analytical solvability or otherwise of the
the resulting models are (i) the spatial structure defined by the
interactions (reflected in which of the $J_{ij}$ are non-zero),
and (ii) the actual values taken by the non-zero interactions
(which will generally be non-trivial, in order to achieve the
objective of the creation of specific attractors).
For the interaction matrix $J_{ij}$ we now make two different choices which both generalise the
model class of \cite{skantzoscoolen}, but in qualitatively different ways.

Our first generalisation focuses on the values of those
interactions which are present, while retaining the mean-field plus
nearest neighbour interaction structure of \cite{skantzoscoolen}:
\begin{equation}
{\rm model\ I:}
\hspace{10mm}
J_{ij}=\sum_{\mu=1}^p\left[\frac{J_\mu^\ell}{N} +
J_\mu^s(\delta_{j,i+1}+\delta_{j,i-1})
\right]\xi_i^\mu\xi_j^\mu
\end{equation}
in which the components $\xi_i^\mu\in\{-1,1\}$ are all drawn
independently at random, with equal probabilities.
Neural networks of this type  correspond to the result of having
stored  in a Hebbian-type fashion a set of $p$ binary patterns
$\bxi_i=(\xi_i^1,\ldots,\xi_i^p)\in\{-1,1\}^N$.  The
neurons can be thought of as arranged on a one dimensional
array with mean-field interactions between all pairs $(i,j)$ given by
$N^{-1}\sum_\mu J_{\mu}^\ell
\xi_i^\mu \xi_j^\mu$, in combination with
interactions between nearest neighbours  neighbours $(i,i+1)$ of strength $\sum_\mu J^s_\mu\xi_i^\mu
\xi_{i+1}^\mu$. We will only consider the case
where $\lim_{N\to\infty}p/ N=0$. The parameters $J_\mu^s$ and
$J_\mu^\ell$ control the embedding strength of pattern $\mu$ in the
short- and long-range synapses,
with negative values corresponding to the creation
of `repellors' rather than attractors. For uniform embedding strengths,
$J_\mu^s=J_s$ and $J_\mu^\ell=J_\ell$ for all $\mu$, we recover
\cite{skantzoscoolen}.

Our second  generalisation affects the spatial structure of the
system, rather than the properties of the attractors (although the latter will be
indirectly affected).
Here our choice of interactions is
\begin{equation}
{\rm model\ II:}
\hspace{10mm}
J_{ij}=\left[\frac{J_\ell}{N} +
J_s^{(1)}(\delta_{j,i+1}+\delta_{j,i-1})+
J_s^{(2)}(\delta_{j,i+2}+\delta_{j,i-2})\right]\sum_{\mu=1}^p\xi_i^\mu\xi_j^\mu
\label{eq:modelII}
\end{equation}
In neural networks of type II, the short-range synaptic
interactions reach beyond nearest-neighbours; here
$J_{\ell},J_s^{(1)},J_s^{(2)}\in\Re$ control the strengths of
long-range, nearest-neighbour and second nearest-neighbour
interactions. Alternatively, in these models the neurons can  be thought of as lying on a
strip, mutually coupled by infinite range interactions of strength
$J_\ell/N\bxi_i\cdot\bxi_j$, in combination with short-range `diagonal'
interactions of strength $J_s^{(1)}\bxi_i\cdot\bxi_{i+1}$ and
`edge' interactions of strength
$J_s^{(2)}\bxi_{i-1}\cdot\bxi_{i+1}$ (see the figure).
Note that the models of type II reduce to those in
\cite{skantzoscoolen} for $J_s^{(2)}=0$.
\begin{figure}[h]
\begin{picture}(200,180)(10,10)
\put(50,50){\line(1,0){350}}
\put(50,150){\line(1,0){350}}
\put(100,50){\circle*{5}}
\put(235,50){\circle*{5}}
\put(368,50){\circle*{5}}
\put(168,150){\circle*{5}}
\put(302,150){\circle*{5}}
\put(100,50){\line(2,3){68}}
\put(168,150){\line(2,-3){68}}
\put(235,50){\line(2,3){68}}
\put(302,150){\line(2,-3){68}}
\put(95,35){$\s_{i-2}$}
\put(230,35){$\s_i$}
\put(365,35){$\s_{i+2}$}
\put(162,160){$\s_{i-1}$}
\put(300,160){$\s_{i+1}$}
\put(140,35){$J_s^{(2)}\bxi_{i-2}\cdot\bxi_i$}
\put(280,35){$J_s^{(2)}\bxi_i\cdot\bxi_{i+2}$}
\put(215,160){$J_s^{(2)}\bxi_{i-1}\cdot\bxi_{i+1}$}
\put(60,100){$J_s^{(1)}\bxi_{i-2}\cdot\bxi_{i-1}$}
\put(347,100){$J_s^{(1)}\bxi_{i+1}\cdot\bxi_{i+2}$}
\put(165,100){$J_s^{(1)}\bxi_{i-1}\cdot\bxi_i$}
\put(255,100){$J_s^{(1)}\bxi_i\cdot\bxi_{i+1}$}
\end{picture}
\label{fig:sketch}
\end{figure}
The most relevant observables in our models are the so-called
overlap order parameters, defined as
$m_\mu(\bs)=N^{-1}\sum_i\xi_i^\mu\s_i$, which measure the degree
of similarity between the actual network state $\bs$ and the
$\mu$-th stored pattern.

Due to the presence of short-range
interactions in the above models (and, similarly, those of \cite{skantzoscoolen}),
the solution of even the simplest scenario where
$p\ll N$ is already significantly more complicated than solving the standard
infinite-range (Hopfield-type, \cite{hopfield,amit}) cases. The solution of both models
will be based on a suitable adaptation of the random-field
techniques of \cite{bruinsma}.

\section{Physics of Model I}

\subsection{Solution via Random Field Techniques}

In order to find the phase diagrams we first isolate the $p$
overlap order parameters by inserting $1=\int
d\bm\,\delta[\bm-\frac1N\sum_{i}\s_i\bxi_i]$ with
$\bm=(m_1,\ldots,m_p)$ and $\bxi_i=(\xi_i^1,\ldots,\xi_i^p)$ in the
expression for the asymptotic free energy per site $f=-\lim_{N\to \infty}(\beta
N)^{-1}\ln Z$, and subsequently replace the delta functions by
their integral representations. For model I this leads to
\[
f=-\lim_{N\to\infty}\frac{1}{\beta N}\ln\int d\bm d\hat{\bm}\
e^{-\beta N\phi_{N}(\bm,\hat{\bm})}
\]
\begin{equation}
\phi_N(\bm,\hat{\bm})=-i\hat{\bm}\cdot\bm -\frac12
\sum_{\mu}J_\mu^\ell m_\mu^2-\frac{1}{\beta N}\ln R_N(\hat{\bm})
\label{eq:energy_modelI}
\end{equation}
where the non-trivial part of the calculation, mainly induced by the
 short-range interactions, has been concentrated  in the term $R_N(\hat{\bm})$
(we consider non-periodic boundary conditions):
\begin{equation}
R_N(\hat{\bm})=\sum_{\bs}\prod_{i=1}^{N-1}
T_{\s_i\s_{i+1}}\
e^{-i\beta \sum_\mu \hat{m}_\mu\s_N\xi _N^\mu},
\hspace{5mm}
T_{\s_i\s_{i+1}}=e^{-i\beta \sum_{\mu}\hat{m}_\mu\s_i\xi_i^\mu+\beta \sum_{\mu}J_\mu^s
(\s_i\xi_{i}^\mu)(\s_{i+1}\xi_{i+1}^\mu)}
\label{eq:R}
\end{equation}
In the limit $N\to\infty$ the above integral will be evaluated via
steepest descent. This results in an expression
for $f$ in terms of the relevant saddle-point of the asymptotic form of $\phi_N$:
$f={\rm extr}_{\bm,\hat{\bm}}\lim_{N\to\infty}\phi_{N}(\bm,\hat{\bm})$.
Since the quantity $R(\hat{\bm})$ does
not contain the order parameters $\bm$, we can immediately take
derivatives in (\ref{eq:energy_modelI}) with respect to $\bm$, which
allows us to eliminate the conjugate variables $\hat{\bm}$ via
$-i\hat{m}_\mu=J_{\mu}^\ell m_\mu$ for all $\mu$. Furthermore we
observe that, since for each $\mu$ the order parameter $m_\mu$ is coupled to
 the infinite-range embedding strengths
$J^\ell_{\mu}$, the so-called `pure state' ansatz $\bm=(m,0,\ldots,0)$ will
automatically render the solution of the model independent of
$J_\mu^\ell$ for all $\mu>1$. From now on we will therefore use the notation
$J^\ell_{1}=J_\ell$. Upon making the pure state ansatz, the resulting simplifications lead to
\begin{equation}
T_{\s_i\s_{i+1}}=e^{\beta J_\ell m\s_i\xi^1_i+\beta \sum_{\mu}J_\mu^s
(\s_i\xi_{i}^\mu)(\s_{i+1}\xi_{i+1}^\mu)}
\end{equation}
To evaluate (\ref{eq:R}) we now first define the quantities
\begin{equation}
R^{(N)}_{\pm}(\bm)
=
\sum_{\bs}\prod_{i=1}^{N-1}T_{\s_i \s_{i+1}}\
e^{\beta J_\ell m\s_N\xi^1_N}\
\delta_{\s_N,\pm 1}
\end{equation}
These allow us to derive a 2$\times$2 stochastic recurrence relation, mapping $\{R^{(N-1)}_\pm\}$
onto $\{R^{(N)}_\pm\}$:
\begin{equation}
\left(\begin{array}{c}
R^{(N)}_+(\bm) \\ R^{(N)}_-(\bm)
\end{array}\right)=
\left(\begin{array}{cc}
e^{\beta(J_\ell m\xi^1_{N}+ \sum_\mu J_\mu^s \xi_{N-1}^\mu \xi_{N}^\mu)} &
e^{\beta(J_\ell m\xi^1_{N} -\sum_\mu J_\mu^s \xi_{N-1}^\mu \xi_{N}^\mu)} \\
e^{-\beta(J_{\ell}m\xi^1_{N}+ \sum_\mu J_\mu^s \xi_{N-1}^\mu \xi_{N}^\mu)} &
e^{-\beta(J_\ell m\xi^1_{N}+ \sum_\mu J_\mu^s \xi_{N-1}^\mu \xi_{N}^\mu)}
\end{array}\right)
\left(\begin{array}{c}
R^{(N-1)}_+(\bm) \\ R^{(N-1)}_-(\bm)
\end{array}\right)
\label{eq:matrix_model1}
\end{equation}
from which the partition sum of (\ref{eq:R}) follows as
\begin{eqnarray}
-\lim_{N\to\infty}\frac{1}{\beta N}\ln R_N(\bm) &=&
-\lim_{N\to\infty}\frac{1}{\beta N}\ln\,[R^{(N)}_+(\bm)
+R^{(N)}_-(\bm)] \nonumber
\\
&=&
-\lim_{N\to\infty}\frac{1}{\beta N}\ln\left\{
\left(\begin{array}{c}
1 \\ 1
\end{array}\right)
\cdot\left[\prod_{i=2}^{N} \bt_i\right]
\left(\begin{array}{c}
R^{(1)}_+(\bm) \\ R^{(1)}_-(\bm)
\end{array}\right)\right\}
\label{eq:energy_model1}
\end{eqnarray}
The successive matrix multiplications above can be simplified via the use of the
following ratio of the conditioned quantities
$\{R^{(j}_+(\bm),R^{(j)}_-(\bm)\}$:
\[
k_{j}=e^{2\beta m J_\ell \xi^1_{j}}\ \frac{R^{(j)}_-(\bm)}{R^{(j)}_+(\bm)}
\]
It now follows  from (\ref{eq:matrix_model1}) that these numbers $k_i$
are, in turn, generated by the following
stochastic process
\begin{equation}
k_{j+1}=\frac{e^{-\sum_{\mu}J_\mu^s \xi_{j}^\mu\xi_{j+1}^\mu}+
k_j\ e^{-\sum_{\mu}J_\mu^s \xi_{j}^\mu\xi_{j+1}^\mu}\ e^{2\beta m J_\ell
\xi_{j}^1}}
{e^{\sum_{\mu}J_\mu^s \xi_{j}^\mu\xi_{j+1}^\mu}+
k_j\ e^{\sum_{\mu}J_\mu^s \xi_{j}^\mu\xi_{j+1}^\mu}\ e^{2\beta m J_\ell \xi_{j}^1}}
\label{eq:processI}
\end{equation}
The stochasticity here is in the pattern components
$\{\xi_i^\mu\}$.
This allows us to work out  the partition sum and express the asymptotic
free energy per neuron as $f={\rm extr}_m f(m)$, with
\begin{equation}
f(m)=\frac12 J_\ell~
m^2-\frac{1}{\beta}\int\!dk\!\!\sum_{\bxi,\bxi^\prime\in\{-1,1\}^p}\rho(k,\bxi,\bxi^\prime)
\log\left\{e^{\beta \sum_\mu J^s_\mu\xi_\mu \xi_\mu^\prime}
+k\ e^{-\beta \sum_\mu J^s_\mu\xi_\mu \xi_\mu^\prime}e^{2\beta
J_{\ell}m\xi_1}\right\}
\label{eq:energy2_modelI}
\end{equation}
with
\begin{equation}
\rho(k,\bxi,\bxi^\prime)
=\lim_{N\to\infty}\frac{1}{N}\sum_{i=1}^{N-1}\delta[k-
k_i]~\delta_{\bxi,\bxi_i}\delta_{\bxi^\prime,\bxi^\prime_{i+1}}
\label{eq:distr_modelI}
\end{equation}
The joint distribution (\ref{eq:distr_modelI}), which is the invariant distribution for the process
(\ref{eq:processI}) and which can be highly
non-trivial \cite{skantzoscoolen} (depending on the choice of system parameters,
the associated integrated density can take the shape of a Devil's Staircase),
is in practice calculated
numerically. In the present case one can in fact simplify matters further
by exploiting symmetry properties of $\rho(k,\bxi,\bxi^\prime)$
resulting from the homogeneous distribution assumed for the
$\{\xi_i^\mu\}$.

\subsection{Phase Diagrams and Comparison with Numerical
Simulations}

\begin{figure}[h]
\vspace*{69mm}
\hbox to
\hsize{\hspace*{3mm}\includegraphics{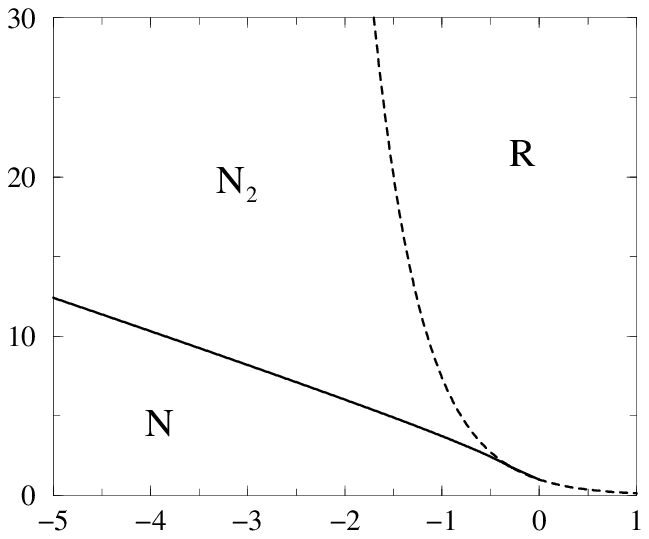}\hspace*{-3mm}}
\vspace*{-4mm}
\hbox to
\hsize{\hspace*{78mm}\includegraphics{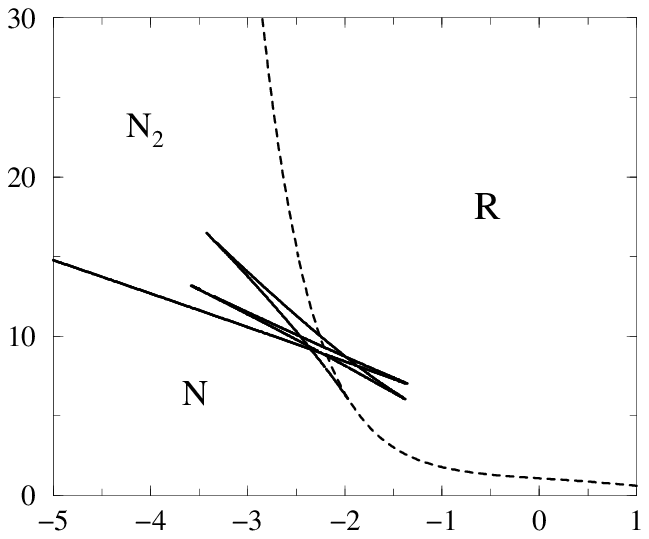}\hspace*{-78mm}}
\vspace*{69mm}
\hbox to
\hsize{\hspace*{3mm}\includegraphics{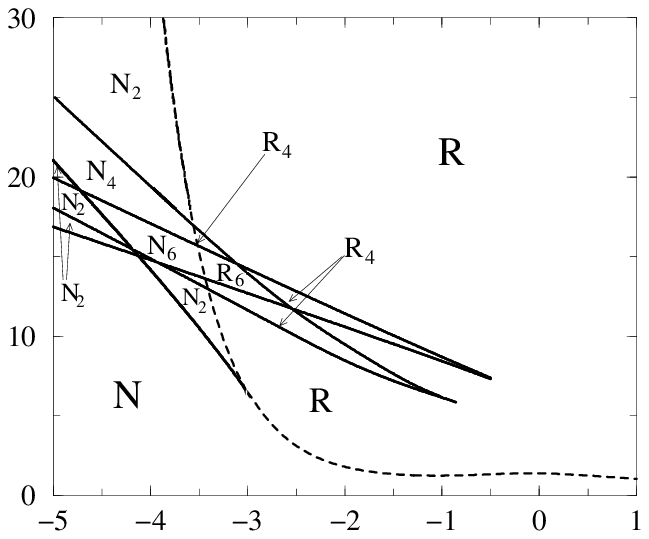}\hspace*{-3mm}}
\vspace*{-4mm}
\hbox to
\hsize{\hspace*{78mm}\includegraphics{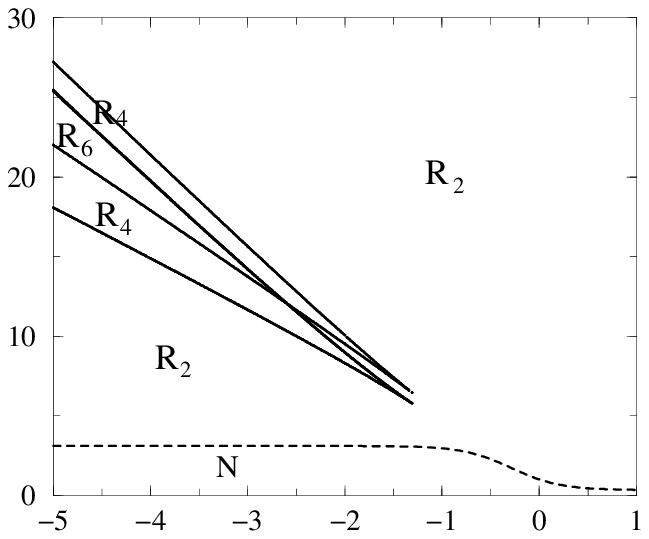}\hspace*{-78mm}}
\vspace*{-32mm}
\begin{picture}(200,85)(10,50)
\hspace*{-5mm}
{\Large
\put(120,445){\mbox{\boldmath{$\beta J^s_2=\!0$}}}
\put(330,445){\mbox{\boldmath{$\beta J^s_2=-1.5$}}}
\put(120,231){\mbox{\boldmath{$\beta J^s_2=-2.5$}}}
\put(330,231){\mbox{\boldmath{$\beta J^s_2=\beta J^s_1$}}}
}
{
\Large
\put(25,360){$\beta J_{\ell}$}
\put(25,150){$\beta J_{\ell}$}
\put(250,150){$\beta J_\ell$}
\put(250,360){$\beta J_\ell$}
\put(150,50){ $\beta J^{s}_1$}
\put(355,50){ $\beta J^s_1$}
\put(150,266){$\beta J^s_1$}
\put(355,266){$\beta J^{s}_1$}
}
\end{picture}
\vspace*{2mm}
\caption{\small
Phase diagram cross-sections of model type I for
$p=2$, upon making the `pure state' ansatz for pattern $\mu=1$,
with $J_\ell=J_1^\ell$.
The parameters $J_1^s$ and $J_2^s$ control
the short-range embedding strength of patterns $\mu=1$ and $\mu=2$,
respectively, $J_\ell$ represents the strength of the mean-field
interactions, and $\beta=T^{-1}$ is the inverse temperature.
 In the absence of pattern $\mu=2$ (upper left) or for
equally strong short-range embedding strengths (lower right), we
recover \cite{skantzoscoolen}.  Solid/dashed lines denote first/second-order
transitions. Regions R and N represent strictly recall or
null-recall regions, whereas R$_i$ and N$_i$ correspond to regions
where the trivial solution $m=0$ is (R) or is not (N) locally stable, and
with $i\in\{2,4,6\}$ giving the number of locally
stable $m\neq 0$ solutions.}
\label{fig:modelI}
\vspace*{-10mm}
\end{figure}

We can now extract the macroscopic characteristics of model I by generating the variables
$\{k_j; \forall j\leq N\}$ numerically (together with the $\{\xi_i^\mu\}$), which leads us to
the joint distribution (\ref{eq:distr_modelI}),  and by subsequently evaluating
(numerically) the local minima of the free energy surface (\ref{eq:energy2_modelI}).
We show in figure \ref{fig:modelI} the resulting phase diagrams,
for $\beta J^s_{2}=\{0,-3/2,-5/2,\beta J^s_{1}\}$ and for the
simplest non-trivial case $p=2$.
In all graphs dashed lines correspond
to second-order transitions and solid lines to first-order ones.

Note that for $J^s_{2}=0$ (see figure \ref{fig:modelI}, upper left panel) only
pattern $\mu=1$ is effectively embedded in the spin chain, and  the phase diagram of
the model is identical to that found earlier in \cite{skantzoscoolen} for $p=1$, as it should.
 In this diagram
we observe, apart from strictly null-recall (N) and recall (R)
phases, that there is also a region in which the trivial solution and two
non-trivial ones (one with positive and one with negative $m$) can be locally
stable simultaneously (indicated by
$\textrm{N}_2$)\footnote{From now on regions which allow for
locally stable
null-recall solutions will be denoted by $\textrm{N}_i$, with $i$
indicating the number of simultaneously locally stable recall solutions.
Similarly, regions which do not allow for locally stable null-recall solutions will be
denoted by $\textrm{R}_i$.}. This region corresponds to parameter
values for which the two different types of synapses compete most
strongly (negative nearest-neighbour interactions versus positive
infinite-range ones). It is separated from the recall region by a
second-order transition (dashed line), and from the null-recall region
by a first-order transition (solid line); these two lines
come together at $\{\beta J_{\ell},\beta
J_1^s\}=\{\sqrt{3},-\frac14\ln3\}$. Another benchmark solution of
\cite{skantzoscoolen} is recovered for the special case of having uniform short-range
embedding strengths $J^s_{2}=J^s_{1}$ (see figure \ref{fig:modelI}, lower right panel).
Here, two `pairs' of first-order transition lines have appeared
which separate the regions $\textrm{R}_4$ and $\textrm{R}_6$ (where
$m=0$ is unstable and where 4 and 6 $m\neq 0$ solutions are
possible, depending on initial conditions) from the N region. Here the second-order
transition line does not touch any of the first-order ones.

The remarkable qualitative difference between  the phase diagrams which the aforementioned two
special cases produce is striking; in our previous study the physical origin
of this difference was not studied. In particular, although the
correctness of the solution had been tested in \cite{skantzoscoolen} against extensive numerical simulations,
 it was not at all clear how and why the second-order transition line (dashed line)
would change from the exponentially rising curve (figure \ref{fig:modelI}, upper left) to the
 one in the lower right corner of figure \ref{fig:modelI}. Our present model
generalizes \cite{skantzoscoolen}, and allows for independent tuning
of the short-range embedding strengths: one can now bring the
non-condensed pattern to life in a continuous way. The top right and lower left
panels
 of figure \ref{fig:modelI}, where $J^s_{2}\neq 0$,
show how the system realises the transition from
the $p=1$ case to the $p=2$ case (where $J_1^s=J_2^s)$. The four panels in the figure can be thought
of as different cross-sections
\vfill
\clearpage
\begin{figure}[t]
\vspace*{60mm}
\hbox to
\hsize{\hspace*{2mm}\includegraphics{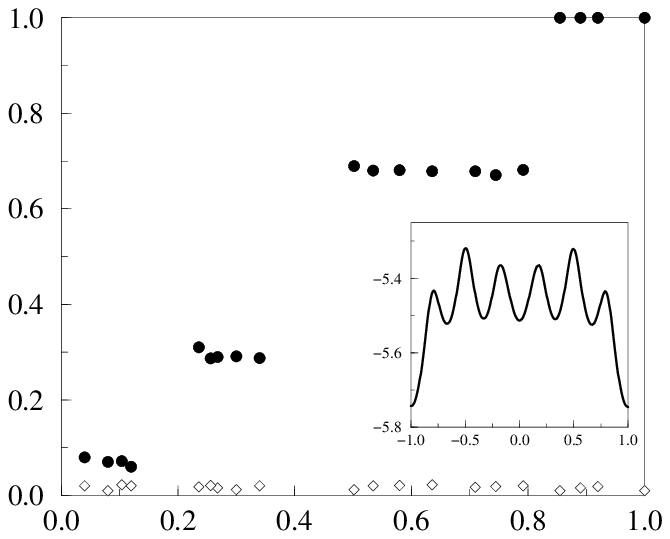}\hspace*{-2mm}}
\vspace*{-5mm}
\hbox to
\hsize{\hspace*{84mm}\includegraphics{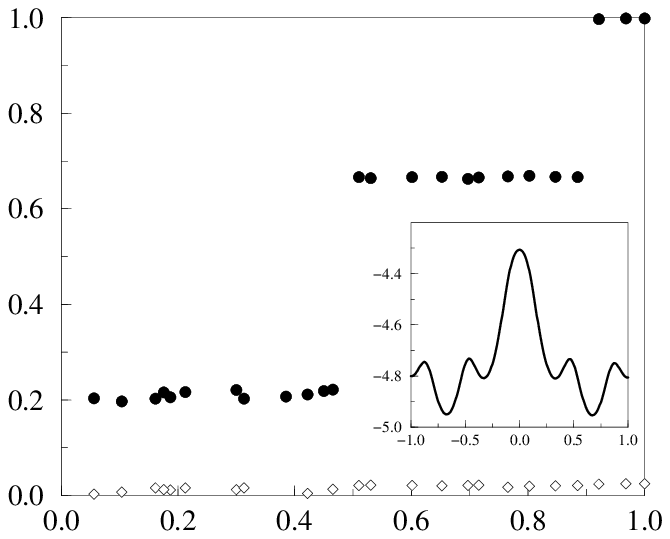}\hspace*{-84mm}}
\begin{picture}(200,0)(10,30)
\hspace*{-5mm}
{
\Large
\put(20,150){$m_{\rm equi}$}
\put(250,150){$m_{\rm equi}$}
\put(140,50){ $m_{\rm init}$}
\put(370,50){ $m_{\rm init}$}
}
\end{picture}
\vspace*{-4mm}
\caption{\small
Simulation results for model type I, for system size $N=1000$ and $p=2$, showing
the equilibrium
value (i.e. that obtained after 10,000 iterations per spin) of the `condensed' overlap $m_{\rm equi}=m_1(t\!\to\!\infty)$ as a function
of the initial value $m_{\rm init}=m_1(t\!=\!0)$ (solid circles).
The initial
configurations were drawn at random, subject to the constraint imposed by the required
value of $m_1(t=0)$. The theoretically predicted locations
of the ergodic components, as contructed from equation
(\ref{eq:energy2_modelI}), are also shown as local minima of the free
energy per neuron in the insets, for comparison. Open diamonds represent the equilibrium
values of the non-condensed overlaps $m_2(t)$; they are seen to remain zero, which
justifies \emph{a posteriori} the `pure state' ansatz. Left picture:
$\beta J^s_2=-3.5,~\beta J^s_1=-4.2,~\beta J_\ell=18$.
Right picture: $\beta J^s_1=-5.5,~\beta J^s_2=-3.5,~\beta J_\ell=23$. }
\label{fig:simulationsI}
\end{figure}

\noindent
of an extended graph in the area $\{\beta
J_1^s, \beta J_2^s, \beta J_\ell\}$, which in combination reveal the
underlying complexity of the model; due to the competing short- and
long-range forces and the high degree of frustration new regions
come to life in parameter space, with multiple locally stable
overlap solutions. In contrast to the $J_1^s=J_2^s$ phase diagram, where
$m=0$ is unstable everywhere, apart from the strictly null-recall
phase, the other two $J_2^s\neq 0$ phase diagrams display
regions ($\textrm{N}_2$, $\textrm{N}_4$ and $\textrm{N}_6$) where
$m=0$ coexists as a locally stable state together with multiple locally  stable $m\neq 0$ solutions.
These latter new phase diagrams appear significantly richer than
those found in \cite{skantzoscoolen}, owing to the breaking of
the pattern embedding strength symmetry.

\begin{figure}[t]
\vspace*{60mm}
\hbox to\hsize{\hspace*{30mm}\includegraphics{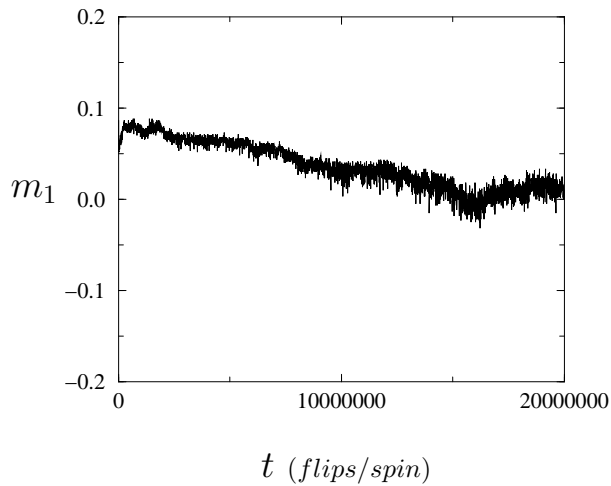}\hspace*{-30mm}}
\begin{picture}(200,0)(10,30)
\hspace*{-5mm}
{
\Large
\put(110,148){$m_1$}
\put(200,42){ $t$ {\small $(flips/spin)$}}
}
\end{picture}
\vspace*{-2mm}
\caption{\small
Simulation results for model type I, for system size $N=1\,200$ and $p=2$, showing
 the `condensed' overlap $m_1$ as a function
of time (measured in iterations per neuron).
The embedding strengths were given by $\beta J^s_2=-3.5,~\beta J^s_1=-4.2,~\beta J_\ell=18$.
The relaxation
towards zero,
following a small (but nonzero) initial value, is seen to be
extremely slow, due to domain formation.}
\label{fig:ageing}
\end{figure}

To test and verify our results we have performed extensive simulation
experiments.  Initial
configurations $\bs(t=0)$ were chosen randomly, according to
\[
p(\bs(0))=\prod_{i}\left\{\frac12
[1+m_0]\delta_{\s_{i}(0),\xi_i^1}+\frac12
[1-m_0]\delta_{\s_{i}(0),-\xi_i^1}\right\}
\]
In figure \ref{fig:simulationsI} we
plot the equilibrium value $m_1(t\to\infty)$ of the main order parameter as a function of
its initial value $m_1(t=0)$ (black circles), in order to probe the existence and location
of multiple ergodic components. To enable comparison with the
theoretically predicted equilibrium values we also show (insets) the
dependence of the asymptotic free energy per neuron on the order parameter $m=m_1(t\to\infty)$,
as contructed from equations
(\ref{eq:energy2_modelI}) and (\ref{eq:energy_modelII}); its local minima are indeed located
at those values which are found as allowed equilibrium states in the simulations,
given appropriate initial conditions, and within the
experimental margin of accuracy.
With our system size $N=1000$, finite size effects are expected to
be of the order of $\mc{O}(N^{-\frac12})\approx 0.03$. Our restriction
to relatively small system sizes was prompted by the appearance of extremely
large equilibration times, due to domain formation.
For the case of predominant long-range interactions
equilibration was achieved within $\approx 10^4$ flips/spin.
For predominant short-range interactions, however,
domain formation led to equilibration times which were
observed to scale exponentially with the system size, see figure
\ref{fig:ageing}.
 For this reason the observed value of $m_{\rm equi}$ for the ergodic component closest
to $m=0$ in figure \ref{fig:simulationsI} appears to differ
from the theoretically predicted value $m=0$ by roughly
$0.06>\mc{O}(N^{-\frac12})$. In figure \ref{fig:ageing} we
show that for the parameter choice of figure
\ref{fig:simulationsI} and for $m_1(t=0)=0.08$ the system is
indeed approaching the predicted state $m_{\rm equil}=0$, but extremely slowly. Finally,
we have also measured the equilibrium overlaps with the
non-`condensed' pattern, $m_2(t\to\infty)$, which are seen to remain
zero (open diamonds in figure \ref{fig:simulationsI}), which justifies our `pure state'
ansatz.

\section{Physics of  Model II}

\subsection{Solution via Random Field techniques}

Neural network models of type II can be solved analytically using the same techniques
as applied to model type I, although here the calculations will be somewhat more elaborate.
Upon again making the `pure state' ansatz: $\bm=(m,0,\ldots,0)$ and upon
eliminating the conjugate
order parameters $\hat{\bm}$ via saddle-point equations, we find that the
asymptotic free energy per neuron is given by $f={\rm extr}_m
f(m)$, with
\[
f(m)=\frac12 J_\ell m^2-\lim_{N\to\infty}\frac{1}{\beta N}\ln R_N(m)
\]
where the complicated part of the partition sum is in the last
term:
\begin{equation}
R_N(m)=\sum_{\bs} \left[\prod_{i=1}^{N-2} T_{\s_i \s_{i+1} \s_{i+2}}\right]
e^{\beta J_s^{(1)}(\s_{N-1}\bxi_{N-1})\cdot(\s_N\bxi_N)}
e^{\beta J_{\ell} m \{\xi_{N-1}\s_{N-1}+\xi_{N}\s_{N}\}}
\label{eq:R_modelII}
\end{equation}
\[
T_{\s_i \s_{i+1} \s_{i+2}}=
e^{\beta
J_s^{(1)}\sum_i(\s_i\bxi_i)\cdot(\s_{i+1}\bxi_{i+1})+\beta
J_s^{(2)}\sum_{i}(\s_i\bxi_i)\cdot(\s_{i+2}\bxi_{i+2}) +\beta
J_\ell m\sum_i\s_i \xi_i}
\]
(with open boundary conditions). As in model I we next derive a recurrence relation
for conditioned partition sums. For the present model we find that
this can be achieved in terms of the following four quantities:
\[
R^{(N)}_{\pm \pm}(\bm)=\sum_{\bs} \prod_{i=1}^{N-2}
T_{\s_i\s_{i+1}\s_{i+2}}\
e^{\beta J_s^{(1)}(\s_{N-1}\bxi_{N-1})\cdot(\s_N\bxi_N)}
e^{\beta J_{\ell} m \{\xi_{N-1}\s_{N-1}+\xi_{N}\s_{N}\}}\
\delta_{\s_{N-1},\pm1 }\ \delta_{\s_{N},\pm1}
\]
These are found to be successively generated by a 4$\times$4 linear but stochastic iterative process
of the form
$\bR_{N+1}=\bt_{N+1} \bR_{N}$, where
$\bR_j=(R^{(j)}_{++},R^{(j)}_{+-},R^{(j)}_{-+},R^{(j)}_{--})$,
the stationary state of which will produce the free energy per
neuron. The 4$\times$4 random matrix $\bt_{N+1}$ can be decomposed further into
two coupled 2$\times$2 random matrices:
\begin{equation}
\left(\begin{array}{c}
R^{(N+1)}_{++} \\ R^{(N+1)}_{+-}
\end{array}\right)
=
\left(\begin{array}{cc}
e^{\beta J_\ell m\xi_{N+1}} & 0\\
0& e^{-\beta J_\ell m\xi_{N+1}}
\end{array}\right)
\left(\begin{array}{cc}
L_{N,+} & L_{N,-} \\
L_{N,+}^{-1} & L^{-1}_{N,-}
\end{array}\right)
\left(\begin{array}{c}
R^{(N)}_{++} \\ R^{(N)}_{-+}
\end{array}\right)
\label{eq:matrix1_model2}
\end{equation}
\begin{equation}
\left(\begin{array}{c}
R^{(N+1)}_{-+} \\ R^{(N+1)}_{--}
\end{array}\right)
=
\left(\begin{array}{cc}
e^{\beta J_\ell m\xi_{N+1}} & 0\\
0& e^{-\beta J_\ell m\xi_{N+1}}
\end{array}\right)
\left(\begin{array}{cc}
L_{N,-}^{-1} & L^{-1}_{N,+} \\
L_{N,-} & L_{N,+}
\end{array}\right)
\left(\begin{array}{c}
R^{(N)}_{+-} \\ R^{(N)}_{--}
\end{array}\right)
\label{eq:matrix2_model2}
\end{equation}
where
\[
L_{N,\pm}=e^{\beta
(J_s^{(1)}\bxi_{N-1}\cdot\bxi_{N}\pm
J_s^{(2)}\bxi_{N-1}\cdot\bxi_{N+1})}
\]
The partition sum in (\ref{eq:R_modelII}) can now be
written in terms of the successive multiplication of the random matrices $\bt$:
\[
-\lim_{N\to\infty}\frac{1}{\beta N}\ln R_N=
-\lim_{N\to\infty}\frac{1}{\beta N}\ln\left\{
\left(\!\!\begin{array}{c}1\\1\\1\\1\end{array}\!\!\right)
\cdot\left[\prod_{i=3}^N\bt_i\right]\bR_2
\right\}
\]
Similar to the analysis performed for model I we
again define ratios of conditioned partition functions (although here we will need three
rather than one):
\[
k_{j}^{(1)}=e^{-2\beta J_{\ell} m\xi_{j}}
\frac{R^{(j)}_{++}}{R^{(j)}_{+-}}
\hspace{10mm}
k_{j}^{(2)}=e^{2\beta J_{\ell} m\xi_{j}}
\frac{R^{(j)}_{+-}}{R^{(j)}_{-+}}
\hspace{10mm}
k_{j}^{(3)}=e^{-2\beta J_{\ell} m\xi_{j}}
\frac{R^{(j)}_{-+}}{R^{(j)}_{--}}
\]
According to (\ref{eq:matrix1_model2}-\ref{eq:matrix2_model2})
these ratios are generated by the following stochastic processes:
\[
k_{j+1}^{(1)}=
\frac{e^{\beta J_s^{(2)}\bxi_{j-1}\cdot\bxi_{j+1}}\
k_{j}^{(1)}k_j^{(2)}+e^{-\beta
J_s^{(2)}\bxi_{j-1}\cdot\bxi_{j+1}}}
{e^{-\beta J_s^{(2)}\bxi_{j-1}\cdot\bxi_{j+1}}\
k_j^{(1)}k_j^{(2)}+e^{\beta J_s^{(2)}\bxi_{j-1}\cdot\bxi_{j+1}}}\
e^{2\beta J_s^{(1)}\bxi_{j-1}\cdot \bxi_{j}}
\]
\[
k_{j+1}^{(2)}=
\frac{e^{-\beta J_s^{(2)}\bxi_{j-1}\cdot\bxi_{j+1}}\
k_{j}^{(1)}k_j^{(2)}+e^{\beta
J_s^{(2)}\bxi_{j-1}\cdot\bxi_{j+1}}}
{e^{\beta J_s^{(2)}\bxi_{j-1}\cdot\bxi_{j+1}}\
k_j^{(2)}k_j^{(3)}+e^{-\beta J_s^{(2)}\bxi_{j-1}\cdot\bxi_{j+1}}}\
k_j^{(3)}\ e^{2\beta J_\ell m \xi_j}
\]
\[
k_{j+1}^{(3)}=
\frac{e^{\beta J_s^{(2)}\bxi_{j-1}\cdot\bxi_{j+1}}\
k_{j}^{(2)}k_j^{(3)}+e^{-\beta
J_s^{(2)}\bxi_{j-1}\cdot\bxi_{j+1}}}
{e^{-\beta J_s^{(2)}\bxi_{j-1}\cdot\bxi_{j+1}}\
k_j^{(2)}k_j^{(3)}+e^{\beta J_s^{(2)}\bxi_{j-1}\cdot\bxi_{j+1}}}\
e^{-2\beta J_s^{(1)}\bxi_{j-1}\cdot \bxi_{j}}
\]
The free energy per neuron can now be expressed in terms of the stationary distribution
of this stochastic process:
\begin{equation}
f(m)=\frac12 J_{\ell}m^2
-\lim_{N\to\infty}\frac{1}{\beta}
\int\!d\bk\!\!\sum_{\bxi,\bxi^\prime\in\{-1,1\}^p}\rho(\bk,\bxi,\bxi^\prime)~
\ln\left\{e^{\beta J_s^{(2)} \bxi\cdot\bxi^\prime}+e^{-\beta J_s^{(2)}
\bxi\cdot\bxi^\prime}\ k^{(2)}k^{(3)}\right\}
\label{eq:energy_modelII}
\end{equation}
with
\begin{equation}
\rho(\bk,\bxi,\bxi^\prime)
=\lim_{N\to\infty}\frac{1}{N}\sum_{i=1}^{N-3}\delta[\bk-\bk_i]~
\delta_{\bxi,\bxi_i}\delta_{\bxi^\prime,\bxi^\prime_{i+2}}
\label{eq:distr_modelII}
\end{equation}
and $\bk_i=(k_i^{(1)},k_i^{(2)},k_i^{(3)})$.

\subsection{Phase Diagrams and Comparison with Numerical
Simulations}

\begin{figure}[t]
\vspace*{69mm}
\hbox to
\hsize{\hspace*{3mm}\includegraphics{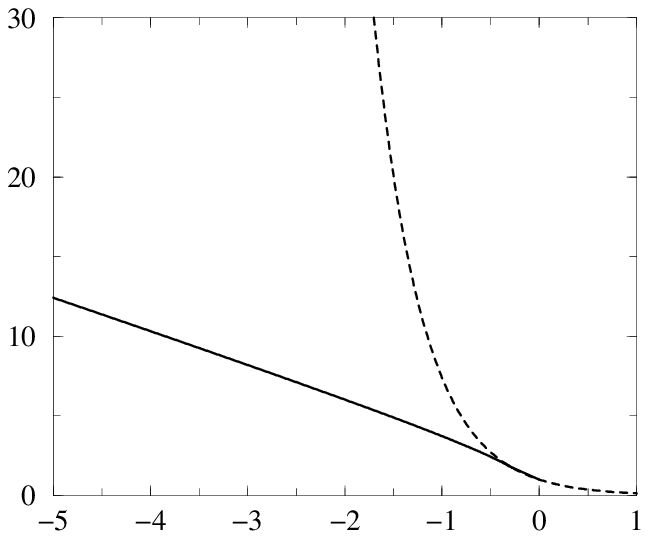}\hspace*{-3mm}}
\vspace*{-4mm}
\hbox to
\hsize{\hspace*{78mm}\includegraphics{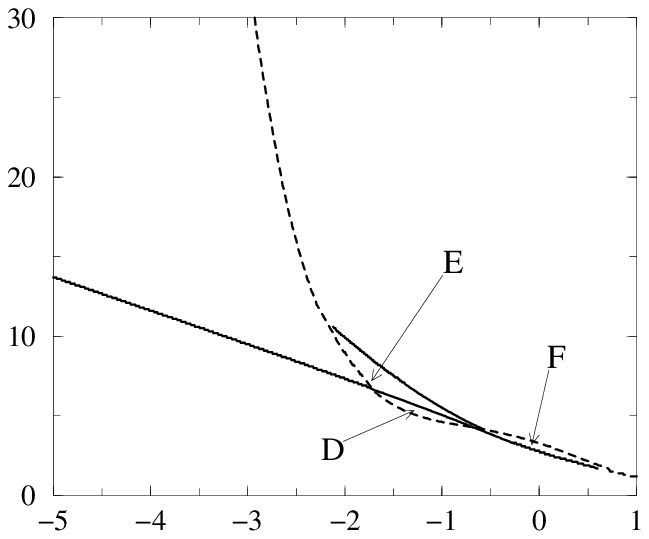}\hspace*{-78mm}}
\vspace*{69mm}
\hbox to
\hsize{\hspace*{3mm}\includegraphics{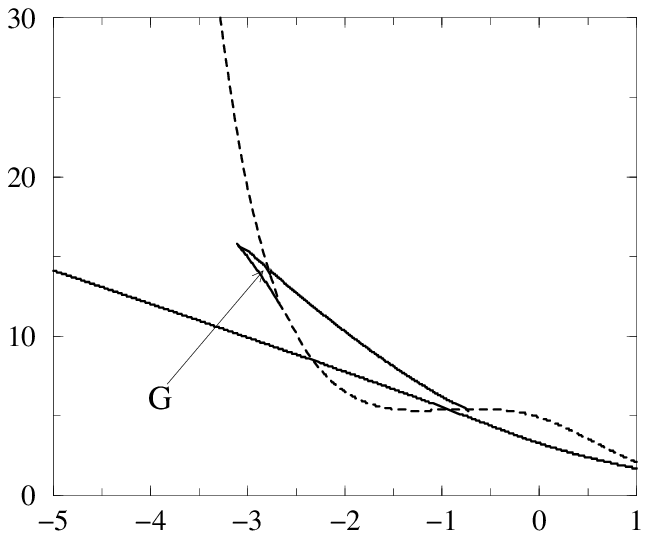}\hspace*{-3mm}}
\vspace*{-4mm}
\hbox to
\hsize{\hspace*{78mm}\includegraphics{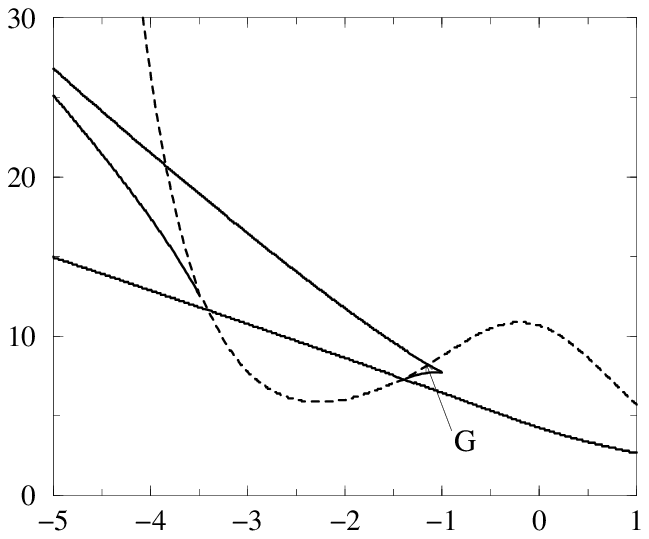}\hspace*{-78mm}}
\vspace*{-32mm}
\begin{picture}(200,85)(10,50)
\hspace*{-5mm}
{\Large
\put(120,445){\mbox{\boldmath{$\beta J_s^{(2)}=\!0$}}}
\put(330,445){\mbox{\boldmath{$\beta J_s^{(2)}=\!-0.6$}}}
\put(120,231){\mbox{\boldmath{$\beta J_s^{(2)}=\!-0.8$}}}
\put(330,231){\mbox{\boldmath{$\beta J_s^{(2)}=\!-1.2$}}}
}
{
\Large
\put(25,360){$\beta J_{\ell}$}
\put(25,150){$\beta J_{\ell}$}
\put(250,150){$\beta J_\ell$}
\put(250,360){$\beta J_\ell$}
\put(150,50){ $\beta J_{s}^{(1)}$}
\put(355,50){ $\beta J_s^{(1)}$}
\put(150,266){$\beta J_s^{(1)}$}
\put(355,266){$\beta J_{s}^{(1)}$}
\put(370,417){\small E: $|m_1|\!>\!|m_2|\!>0$}
\put(370,405){\small F: $m\!=\!0\ \&\ m\!\neq\! 0$}
\put(370,393){\small D: $m\!\neq\! 0\ only$}
\put(160,206){\small G: $|m_1|\!>\!|m_2|\!>\!0$}
\put(175,191){\small  $\&\ m\!=\!0 $}
\put(95,375){\small $m\!=\!0\ \&\ m\!\neq\! 0$}
\put(100,310){\small $m\!=\!0$}
\put(195,380){\small $m\!\neq\! 0$}
}
\end{picture}
\vspace*{2mm}
\caption{\small
Phase diagram cross-sections  of model type II for $p=1$, with
$J_s^{(1)},~J_s^{(2)}$ and $J_\ell$ denoting nearest-neighbour,
next-nearest neighbour, and long-range embedding strengths, respectively. In
the absence of next-nearest neighbour interactions (upper left)
we recover the model of \cite{skantzoscoolen}, whereas for $J_s^{(2)}\neq 0$ new
regions appear in the phase diagram, with different numbers of
simultaneously locally stable solutions for the `overlap' order parameter $m$.
Solid lines denote
first-order transitions,  dashed lines denote second-order
ones.}
\label{fig:modelII_p1}
\end{figure}

\begin{figure}[t]
\vspace*{72mm}
\hbox to
\hsize{\hspace*{2mm}\includegraphics{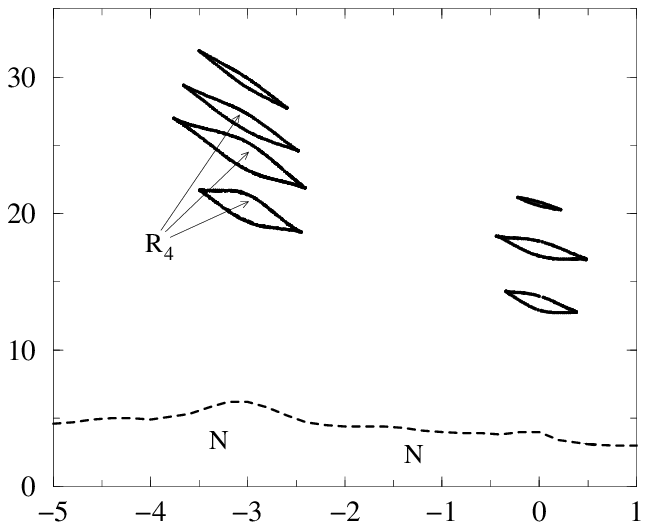}\hspace*{-2mm}}
\vspace*{-4.8mm}
\hbox to
\hsize{\hspace*{80mm}\includegraphics{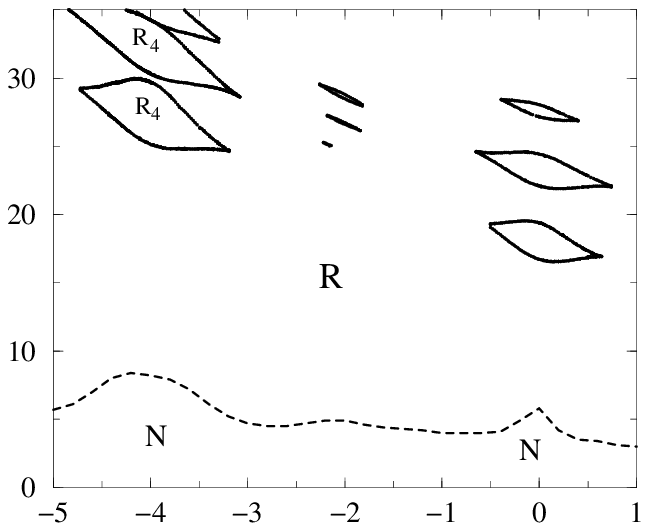}\hspace*{-80mm}}
\begin{picture}(200,85)(10,50)
\hspace*{-5mm}
{\Large
\put(120,335){\mbox{\boldmath{$\beta J_s^{(2)}=\!-3$}}}
\put(330,335){\mbox{\boldmath{$\beta J_s^{(2)}=\!-4$}}}
}
{
\Large
\put(25,240){$\beta J_{\ell}$}
\put(250,240){$\beta J_\ell$}
\put(145,140){ $\beta J_{s}^{(1)}$}
\put(355,140){$\beta J_s^{(1)}$}
}
\end{picture}
\vspace*{-30mm}
\caption{\small
Phase diagram cross-sections for model type II, for $p=5$ and upon making the
`pure state' ansatz, with
$J_s^{(1)},J_s^{(2)}$ and $J_\ell$ denoting nearest-neighbour,
next-nearest neighbour, and long-range embedding strengths, respectively.
The two graphs show typical results for the $p>1$ phase phenomenology.
The `islands' correspond to regions with
four simultaneously locally stable states (two with positive $m$, and two with negative
$m$). The structural differences between the $p=1$ and $p>1$ diagrams can be understood
upon modulating the embedding strengths of the stored patterns, as with model type I.
}
\label{fig:modelII_p5}
\end{figure}

Numerical evaluation of the energy surface defined by
(\ref{eq:energy_modelII}) leads to the phase diagrams shown in figures
\ref{fig:modelII_p1} and \ref{fig:modelII_p5}, which describe the
cases $p=1$ and $p=5$, respectively. They are drawn in the $\{\beta
J_s^{(1)},\beta J_\ell\}$ plane, for four different values of
the next-nearest neighbour embedding strength $J_s^{(2)}$.
In all
phase diagrams the solid lines represent continuous (second-order) phase transitions,
whereas the dashed lines correspond to discontinuous (first-order) ones.
In the absence of next-nearest second neighbour interactions, i.e. for
$J_s^{(2)}=0$ (upper left graph in figure \ref{fig:modelII_p1})  our
model reduces to that of \cite{skantzoscoolen}.
For $J_s^{(2)}>0$ one finds  no new phase
regimes, compared to the $J_s^{(2)}=0$ case; the two transition lines
of the $J_s^{(2)}=0$ phase diagram are found to simply move towards $\beta
J_s^{(1)}=\infty$. However, as soon as $J_s^{(2)}<0$, frustration
effects become more important, with new
regions appearing in
the phase diagram as a result. In the upper right graph of figure \ref{fig:modelII_p1},
 where $\beta J_s^{(2)}=-0.6$, we observe that three new regions have been
created: region \textbf{D} (with $m\neq 0$), region \textbf{F} (where $m=0$
and $m\neq 0$ are simultaneously locally stable) and
region \textbf{E} (with two positive and two negative locally
stable $m\neq 0$ states). In the lower left graph, where
$\beta J_s^{(2)}=-0.8$, we see that, in addition to the previously
created regions, a further new region \textbf{G} comes to life, where the trivial state
as well as four $m\neq 0$ ones are all simultaneously locally stable.
 For $p>1$,
first-order transition lines are found to emerge as boundaries of `islands' in the $\{\beta
J_{s}^{(1)},\beta J_{\ell}\}$ plane, where four $m\neq 0$ solutions
are simultaneously locally stable. For increasingly negative values of
$J_s^{(2)}$ these islands
expand in size, and at some point start overlapping, which creates additional new regions. In
figure \ref{fig:modelII_p5} we show typical phase diagrams for the
case $p=5$. Extensive numerical work also shows that
increasing the number of patterns $p$ further leads to the appearance of
further transition lines in the phase diagrams. This is due to the
explicit dependence of the free energy per neuron, as defined  by equation
(\ref{eq:energy_modelII}), on $p$. Unlike the more conventional long-range  Hopfield-type
networks \cite{hopfield,amit}, where after the `pure state' ansatz has been made the
macroscopic observables have become independent of $p$, in the present model
the short-range interactions ensure that thermal fluctuations around a pure state
will always induce no-negligible $p$-dependent interference
on the recall of the pure state.

\begin{figure}[b]
\vspace*{65mm}
\hbox to
\hsize{\hspace*{2mm}\includegraphics{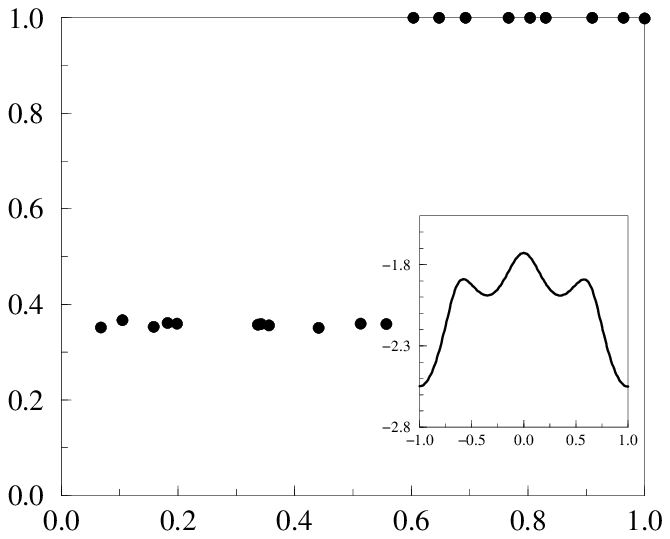}\hspace*{-2mm}}
\vspace*{-5mm}
\hbox to
\hsize{\hspace*{84mm}\includegraphics{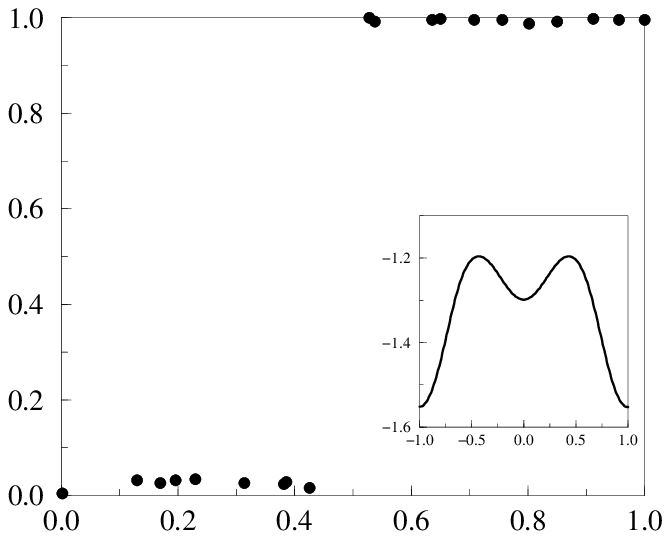}\hspace*{-84mm}}
\begin{picture}(200,0)(10,30)
\hspace*{-5mm}
{
\Large
\put(20,150){$m_{\rm equi}$}
\put(250,150){$m_{\rm equi}$}
\put(140,50){ $m_{\rm init}$}
\put(370,50){ $m_{\rm init}$}
}
\end{picture}
\vspace*{-4mm}
\caption{\small
Simulation results for model II, for system size $N=1\,000$ and $p=1$, 
showing the equilibrium
value of the `condensed' overlap 
$m_{\rm equi}=m_1(t\!\to\!\infty)$ as a function
of the initial value $m_{\rm init}=m_1(t\!=\!0)$ (solid circles). The initial
configurations were drawn at random, subject to the constraint
imposed by the required value of $m_1(t\!=\!0)$.
The theoretically predicted locations
of the ergodic components, as constructed from equation
(\ref{eq:energy_modelII}),
are also shown as local minima of the free
energy per neuron in the insets, for comparison.
Left picture: $\beta J_s^{(1)}=-2.5,~\beta J_s^{(2)}=-1.2,~\beta J_\ell=12.5$.
Right picture: $\beta J_s^{(1)}=-.5,~\beta J_s^{(2)}=-1.2,~\beta J_{\ell}=6.5$. }
\label{fig:simulationsII}
\end{figure}

To test our results we have again performed extensive numerical simulation
experiments, the  results of which are shown in figure \ref{fig:simulationsII}.
The initial states were drawn similar to those in the simulations
of model type I.
We plot the equilibrium overlap oder parameter $m_1(t\to\infty)$ as a function of
its initial value $m_1(t=0)$, to probe different ergodic components,
and compare the locations of these components (in terms of the associated values of $m_1(t\to\infty)$)
with the
\clearpage

\clearpage

\noindent
theoretical prediction by also showing (inset graphs) the
asymptotic free energy per site as constructed from equations
(\ref{eq:energy2_modelI}) and (\ref{eq:energy_modelII}).
The agreement between the two is quite satisfactory,
 and well
within the error margin given by the
finite size effects (with $N=1000$ these are estimated at $\mc{O}(N^{-\frac12})\approx 0.03$).

\section{Discussion}

In this paper we presented an equilibrium statistical mechanical
analysis of a generalized family of recurrent neural
network models, with information stored in the form of attractors in the neuronal state space,
but with 1D spatial structure and competition between
short-range and long-range information processing.
We have solved two specific classes of problems. In the first class, patterns
are embedded in both the long-range and nearest neighbour interactions of the
neuronal chain, but with pattern-dependent embedding strengths (similar
to \cite{viana}). This generalizes a 
previous study \cite{skantzoscoolen}, where all
embedding strengths were independent of the pattern labels.
The breaking of the previous embedding strength symmetry is found to
yield significantly richer phase diagrams, and, moreover, serves
to elucidate the remarkable structural 
differences which were observed (but not understood)
in \cite{skantzoscoolen}
between the phase diagrams for the two simplest
cases $p=1$ and $p=2$. In our second class of models, 
which is a qualitatively different
generalization of the models in \cite{skantzoscoolen},
our neurons are equipped with next-nearest neighbour interactions
(in addition to the long-range and the nearest-neighbour ones),
which increases significantly the potential for 
frustration and competition, given
appropriate choices of the various pattern embedding strengths.
We have been able to solve our models exactly, dealing with the 
random transfer matrix multiplications
in the relevant partition sums (generated by the short-range
interactions)
using suitable adaptations of the random-field techniques 
presented in \cite{bruinsma}.
Alternatively, one could solve our present models
using the random-field
techniques of \cite{rujan}, which provides an 
independent theoretical test of our
solution: we have carried out this test, 
and found full agreement (see appendix).
For both model types we found surprisingly rich phase diagrams, with
qualitatively distinct topologies, and interesting
scenarios of phase diagram deformation when appropriate control
parameters are varied.
Extensive numerical simulation experiments support our theoretical results
convincingly, in terms of the appearance and
location of the multiple ergodic components in phase space.

Note that we have concentrated in this paper on the analysis of the
phenomenology of dynamical phase transitions, i.e.
we have concerned ourselves with the {\em local stability} of
extremal points of the free energy, written as a function of the
main order parameter, rather than with the actual value of the free
energy in the various locally stable states.
In the case of large recurrent neural networks one is simply not
interested in thermodynamic transitions, since the
time-scales relevant to their operation as associative memories
are much smaller than the escape times of locally stable states (which diverge with the
system size); the locally stable states, and their domains of
attraction, determine the relevant physical properties.
The models and methods in this paper can be adapted in a straightforward manner
to cover systems with non-binary neuron states, or
other types of one-dimensional architectures;
the inclusion of more distant short-range interactions, however, will lead to
higher dimensional random transfer matrices, and the calculations will become
more involved. Alternatively, one could turn to
models with synchronous rather than  sequential dynamics.

Long-range models, as in \cite{hopfield,amit},
have been of immense value in shaping our understanding of
information processing in attractor neural networks, but are far
removed from biological reality.
Our present study emphasizes once more the richness of attractor
neural networks with spatial structure, and their analytical
solvability (at least, within the context of $1+\infty$
dimensional models).
Not only can exact solutions be obtained beyond the familiar infinite range models,
but one can also generalize the relatively simple but solvable $1+\infty$ dimensional models of
\cite{skantzoscoolen}, increasing again (albeit with small steps) their biological
relevance. This allows one to investigate further
(quantitatively) the significant impact of simple forms of spatial structure on
information processing via the manipulation of attractors in recurrent neural
networks.

\clearpage
\appendix

\section{Comparison with Rujan's Solution}

An alternative method to calculate the partition sums (\ref{eq:R}) and
(\ref{eq:R_modelII}) is provided by the random-field technique of
\cite{rujan}. This requires the result of each of the individual spin summations to be written in the form
\begin{equation}
{\rm model\ I:}
\hspace{10mm}
\sum_{\s_{j-1}}
T_{\s_{j-1}\s_j}=e^{h_{j}(\xi_{j-1}^\mu,\xi_{j}^\mu)\s_{j}+
L_{j}(\xi_{j-1}^\mu,\xi_{j}^\mu)}
\label{eq:sum_s1}
\end{equation}
for some $\{h_j,L_j\}$, for
$\s_j\in\{-1,1\}$. This transformation allows us to evaluate for
$N\to\infty$ the non-trivial part of (\ref{eq:energy_modelI})
\[
-\lim_{N\to\infty}\frac{1}{\beta
N}\ln R(\bm)=
\]
\[
-\lim_{N\to\infty}\frac{1}{2\beta N}\sum_i
\ln\left[4
\cosh[\beta\sum_\mu (J_\mu^s\xi_i^\mu\xi_{i+1}^\mu)+\beta J_\ell m\xi_i^1+h_i]
\cosh[\beta\sum_\mu (J_\mu^s\xi_i^\mu\xi_{i+1}^\mu)+\beta J_\ell m\xi_i^1-h_i]
\right]
\]
where in the thermodynamic limit we will assume that the
asymptotic distribution of the stochastic variables $\{h_i\}$ is uniquely generated by the
process
\begin{equation}
h_{i+1}=\frac12\ln\frac{
\cosh[\beta\sum_\mu(J_\mu^s\xi_i^\mu\xi_{i+1}^\mu)+\beta J_\ell m\xi_i^1+h_i]}
{\cosh[\beta\sum_\mu(J_\mu^s\xi_i^\mu\xi_{i+1}^\mu)+\beta J_\ell m\xi_i^1-h_i]}
\label{eq:h_map}
\end{equation}
becomes stationary.

In a similar fashion one can perform for $N\to\infty$ the partition
sum in (\ref{eq:R_modelII}) requiring
\begin{equation}
{\rm model\ II:}
\hspace{10mm}
\sum_{\s_{j-1}}
T_{\s_{j-1}\s_j\s_{j+1}}=
e^{h^{(1)}\s_j\s_{j+1}+h^{(2)}\s_j+h^{(3)}\s_{j+1}+L_{j-1}}
\end{equation}
to be true for $\s_j,\s_{j+1}\in\{-1,1\}$. This allows us to write
\[
-\lim_{N\to\infty}\frac{1}{\beta N}\ln R(m)=-\lim_{N\to\infty}\frac{1}{4\beta N}\sum_{i}\ln\left\{
2^4\ \Omega_{++}^{(i)}\ \Omega_{+-}^{(i)}\ \Omega_{-+}^{(i)}\
\Omega_{--}^{(i)}\right\}
\]
where the quantities $\Omega_{\lambda\lambda'}^{(i)}$ are
obtained iteratively as functions of three stochastically evolving
variables $\{h^{(1)}_{j-1},h^{(2)}_{j-1},h^{(3)}_{j-2};\forall\ j\leq i\}$:
\[
\Omega^{(i)}_{\lambda\lambda'}=
\cosh[(\beta
J_s^{(1)}\bxi_i\cdot\bxi_{i+1}+h^{(1)}_{i-1}\theta(i-2))\lambda+\beta
J_s^{(2)}\bxi_{i}\cdot\bxi_{i+2}\lambda'+\beta
J_\ell m\xi_i +h^{(2)}_{i-1}\theta(i-2)+h^{(3)}_{i-2}\theta(i-3)]
\]
where $\theta(j)=1$ if $j\geq 0$ and $\theta(j)=0$ otherwise, and
with
\begin{equation}
h^{(1)}_{i}=\frac14 \ln
\frac{\Omega^{(i)}_{++}\ \Omega_{--}^{(i)}}{\Omega_{+-}^{(i)}\
\Omega_{-+}^{(i)}}
\hspace{10mm}
h^{(2)}_{i}=\frac14 \ln
\frac{\Omega^{(i)}_{++}\ \Omega_{+-}^{(i)}}{\Omega_{-+}^{(i)}\
\Omega_{--}^{(i)}}
\hspace{10mm}
h^{(3)}_{i}=\frac14 \ln
\frac{\Omega^{(i)}_{++}\ \Omega_{-+}^{(i)}}{\Omega_{+-}^{(i)}\
\Omega_{--}^{(i)}}
\label{eq:process2}
\end{equation}
Numerical iteration of the processes (\ref{eq:h_map}) and
(\ref{eq:process2}) and subsequently evaluation of the asymptotic
free energies and of the order parameters shows excellent agreement
with the results found earlier for models I and II.

\end{document}